\documentclass[twocolumn,aps,prl,showpacs]{revtex4-1}

\usepackage{latexsym,amssymb,float}
\usepackage{setspace}
\usepackage{graphicx}
\usepackage{epsfig}
\usepackage{amsmath}
\usepackage{bm}

\begin{document}

\title{Atomic Resolution Imaging of Currents in Nanoscopic Quantum Networks via Scanning Tunneling Microscopy}

\author{Tankut Can$^{1}$ and Dirk K. Morr$^{1,2}$}
\affiliation{$^{1}$ Department of Physics and James Franck
Institute, University of Chicago, Chicago, IL 60637, USA \\
$^{2}$University of Illinois at Chicago, Chicago, IL 60607, USA}
\date{\today}
\begin{abstract}
We propose a new method for atomic-scale imaging of spatial current patterns in nanoscopic quantum networks by using scanning tunneling microscopy (STM). By measuring the current flowing from the STM tip into one of the leads attached to the network as a function of tip position, one obtains an atomically resolved spatial image of ``current riverbeds" whose spatial structure reflects the coherent flow of electrons out of equilibrium. We show that this method can be successfully applied in variety of network topologies, and is robust against dephasing effects.

\end{abstract}


\maketitle

How does a current flow through a nanoscopic system? The answer to this question is not only of fundamental interest for our understanding of quantum phenomena at the nanoscale, but also of great importance for the further development of nano-electronics and the continued miniaturization of electronic devices. In mesoscopic systems, such as quantum point contacts \cite{Cro00,Top00,Jura07}, quantum rings \cite{Hac06} and DNA \cite{Ter05}, this question has been successfully answered by imaging spatial current paths using a scanning probe microscope (SPM) \cite{Cro00,Top00,Ter05,Hac06,Jura07,reviews, Met05}. However, in nanoscopic systems with sizes of tens of nanometers, this question has only been addressed theoretically so far \cite{CLoops,Tod99,Can12}. The main experimental challenge here arises from the SPM's perturbing electrostatic potential, which yields a spatial resolution that is insufficient to detect the atomic-scale varying current patterns predicted to exist in nanoscopic systems \cite{CLoops,Tod99,Can12}. Developing an experimental method that allows atomically resolved imaging of spatial current paths is therefore of great importance for understanding charge transport at the nanoscale, opening new possibilities to explore quantum Hall physics \cite{Aoki05} or weak localization effects \cite{Ber10} in nanoscopic systems.

In this Letter, we propose such a method by demonstrating that atomic-scale imaging of spatial current paths in nanoscopic quantum networks can be achieved using scanning tunneling microscopy (STM) \cite{Hof03}.
In particular, by measuring the current flowing from an STM tip into one of the leads attached to the network as a function of tip position, one reveals atomically resolved ``current riverbeds" whose spatial structure reflects the coherent flow of electrons out of equilibrium. We show that this method can be successfully applied in a variety of network topologies, and  reproduces even complex current patterns arising from the presence of constrictions. It is also robust against dephasing effects, providing correct current images even in systems where the mean-free path is only a few lattice constants. Finally, we demonstrate that the form of the network's conductance is an experimentally accessible criterion for the success of the imaging method. Thus, the proposed STM imaging method  provides a promising new approach for exploring charge transport in nanoscopic systems.

We study the spatial current patterns in a nanoscale quantum network that is attached to two leads and described by the Hamiltonian
\begin{eqnarray}
{\cal H}_c & = &  -  t \sum_{{\bf r},{\bf
r}^\prime,\sigma} \; c^\dagger_{{\bf r},\sigma} c_{{\bf r}^\prime,\sigma} - t_{l} \sum_{{\bf l},{\bf l}^{\prime},\sigma} \; d^\dagger_{{\bf l},\sigma} d_{{\bf l}', \sigma}  \nonumber \\
 & & - t_h \sum_{{\bf r},{\bf l},\sigma} \; \left( c^\dagger_{{\bf r},\sigma} d_{{\bf l}, \sigma} + d^\dagger_{{\bf l},\sigma} c_{{\bf r}, \sigma} \right) \nonumber \\
 & & \nonumber \\
& & + \sum_{\bf r} \omega_0 \ a_{\bf r}^\dagger a_{\bf r} + g \sum_{\bf r, \sigma} (a_{\bf r}^\dagger + a_{\bf r}) c^\dagger_{{\bf r},\sigma} c_{{\bf r},\sigma} \ .
\label{eq:Hamiltonian}
\end{eqnarray}
Here, $c^\dagger_{{\bf r},\sigma}$ and $d^\dagger_{{\bf r}, \sigma}$ create an electron with spin $\sigma$ at site
{\bf r} in the network and leads, respectively, with $t, t_{l}$ and $t_h$ being the respective hopping elements. While the network sites can in general represent atoms, molecules or more complex structures, we assume here that each site contains only a single relevant electronic level. The last two terms describe a local electron-phonon interaction in the network with a phonon mode of energy $\omega_0$.

\begin{figure}
\includegraphics[width=8.cm]{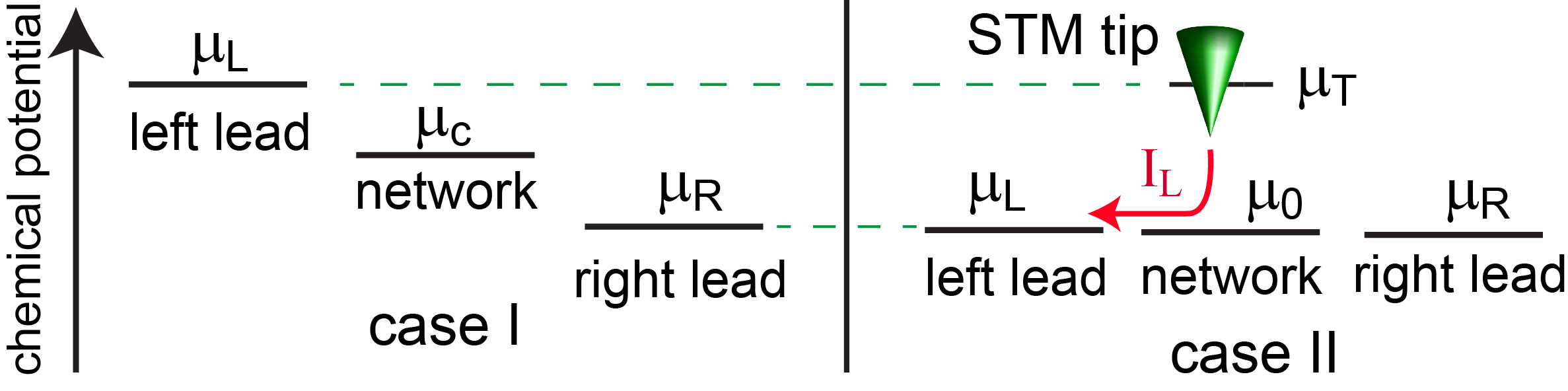}
\caption{(color online) Position of the chemical potentials of the leads, the STM tip, and the network for cases I and II. } \label{fig:Fig1b}
\end{figure}
For different chemical potentials, $\mu_{L,R}$ in the left and right leads (see Fig.~\ref{fig:Fig1b}, case I), a spatial current pattern is obtained by computing the current, $I_{{\bf r},{\bf r}^\prime}$, between adjacent sites ${\bf r}$,${\bf r}^\prime$ in the network via \cite{CurrentTheory, Car71a}
\begin{equation}
I_{{\bf r},{\bf r}^\prime}=-2 \frac{e}{\hbar} \;
\intop_{-\infty}^{+\infty}\frac{d\omega}{2\pi} t_{{\bf r},{\bf r}^\prime} {\rm Re} \left[{\hat
G}^K_{{\bf r},{\bf r}^\prime}(\omega)\right] \ , \label{eq:Current}
\end{equation}
where, ${\hat G}^{K}$ is the full Keldysh Green's function matrix of the entire system. For a non-interacting ($g=0$) network, one has
$ \hat{G}^{K} =
\left(1-\hat{g}^{r}\hat{t}\right)^{-1}\hat{g}^{K}\left(1-\hat{t}\hat{g}^{a}\right)^{-1}$
with $
{\hat g}^{K}(\omega) = 2i \left[1-2 {\hat n_F}(\omega) \right] {\rm Im}
\left[ {\hat g}^r(\omega) \right]$.  Here, $\hat{t}$ is the symmetric hopping matrix and the matrices $\hat{g}^{r,a,K}$ contain the decoupled ($t,t_h=0$) retarded,
advanced and Keldysh Green's functions of the  network sites [with $g^{r}=1/(\omega +i\delta$), $\delta=0^+$] and leads.
The latter are computed using a renormalization procedure \cite{Gro89}. ${\hat n_F}$ contains the
Fermi functions of the network and leads. Below, we use for concreteness $t=t_h=t_l$, $\delta = 0.0002t, \Delta \mu =\mu_L-\mu_R= 10^{-4}t$, and $k_{B} T = 10^{-5} t$.
In the limit
$\Delta V =\Delta \mu /e \rightarrow 0$ and temperature $T=0$, one obtains from Eq.(\ref{eq:Current}) \cite{Can12}
\begin{equation}
I_{{\bf r, r}^\prime}=4 N_{L} t_h^2 \frac{t e^{2}}{\hbar} {\rm Im}
\left[G^{r}_{{\bf r,L}} G^{a}_{{\bf L,r^\prime}}\right] |_{\omega=\mu_c}
\Delta V \label{eq:Ianalytic}
\end{equation}
where $\mu_c=(\mu_L+\mu_R)/2$ and $N_{L}$ is the density of states (DOS) in the left lead.

To image the spatial current patterns obtained from Eq.(\ref{eq:Current}), we compute the current, $I_{L,R}({\bf T})$, flowing from an STM tip (held at potential $\mu_{T}$) into the left (L) or right (R) lead as a function of tip position ${\bf T}$ (see Fig.~\ref{fig:Fig1b}, case II).
Here, ${\cal H}_T = - t_T \sum_{{\bf r},\sigma} \; \left( c^\dagger_{{\bf r},\sigma} f_{\sigma} + H.c.\right)$ describes the electron tunneling between the STM tip and a single network site with $f^\dagger$ creating an electron in the tip, and one appropriately extends ${\hat G}^{K}$ in Eq.(\ref{eq:Current}).
To probe the same states participating in the charge transport as in case I, we set the bias between the leads to zero, i.e., $\mu_{L,R} = \mu_{0}$, and require $\Delta V_T = (\mu_T - \mu_0) /e = \Delta V$ and $\mu^T_c = (\mu_0 + \mu_T)/2= \mu_c$ with $\Delta V$ and $\mu_c$ from case I (see Fig.~\ref{fig:Fig1b}). Assuming, for example, that the left lead is connected to a single network site ${\bf L}$ only, one obtains for $\Delta V_T \rightarrow 0$ and $T=0$
\begin{equation}
I_{L}({\bf T}) = 2 \frac{e^{2} \pi}{ \hbar} t_T^2 t_{h}^2 N_{L} N_T  \left|G^r_{{\bf T,L}}\right|_{\omega=\mu_c^T}^{2}\Delta V_T \ ,
\label{eq:IL}
\end{equation}
where $N_T$ is the DOS in the STM tip. Note that in contrast to $I_{L}$, the total tip current, $I_{tip}=I_{L} + I_{R}$, for $\Delta V_{T} \rightarrow 0$ yields a spatial image of the network's density of states at $\omega = \mu_{c}^{T}$. In the experimentally realized weak tunneling limit \cite{Hof03}, the STM only probes but essentially does not perturb the network's electronic structure, and is thus less intrusive than the SPM imaging method \cite{Cro00,Top00,Ter05,Hac06,Jura07,reviews, Met05}.

\begin{figure}
\includegraphics[width=8.cm]{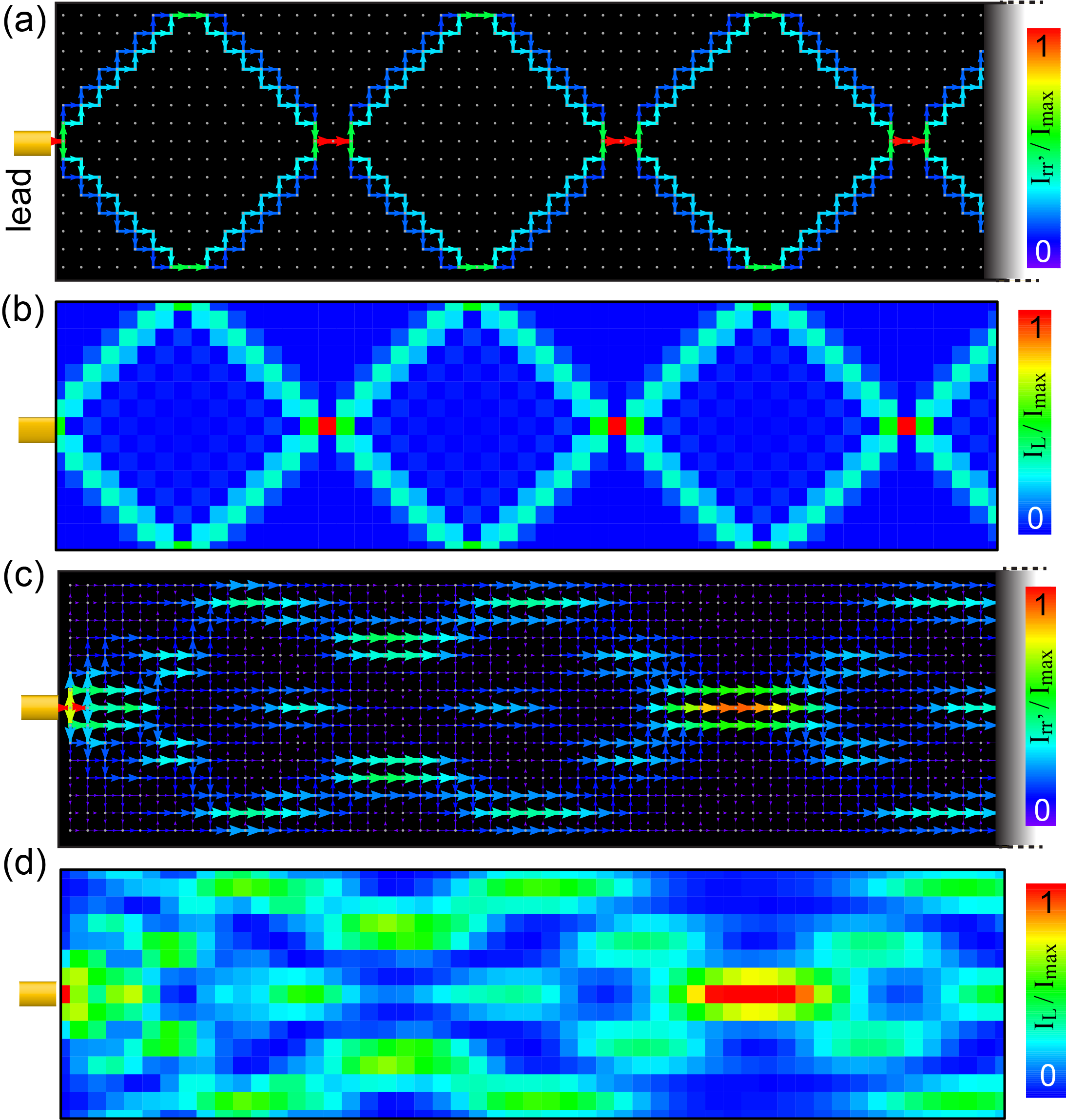}
\caption{(color online) (a) Spatial current pattern, $I_{{\bf r},{\bf r}^\prime}$, and (b) $I_L({\bf T})$ for $\mu_c=0$. (c) $I_{{\bf r},{\bf r}^\prime}$, and (d) $I_L({\bf T})$ for $\mu_c=2.4t$. All currents are normalized.} \label{fig:Fig1}
\end{figure}

We begin by considering the spatial current pattern, $I_{{\bf r},{\bf r}^\prime}$, obtained from Eq.(\ref{eq:Current}) (case I) in a non-interacting, semi-infinite network with $N_y=15$ rows that is connected to a narrow left lead through a single site, ${\bf L}$ (see Fig.~\ref{fig:Fig1}). For $\mu_c=0$ [Fig.~\ref{fig:Fig1}(a)], the Fermi velocity of the states involved in the current transport points along the lattice diagonal, resulting in a diagonal current path that bounces off the sides of the quantum network \cite{Can12}.
A comparison of this current pattern with $I_L$ obtained from Eq.(\ref{eq:IL}) (case II) shown in Fig.~\ref{fig:Fig1}(b) demonstrates that $I_L$ provides an atomically resolved image of the current paths in the network, a result which also holds when the network is gated, and $\mu_c$ is shifted, as exemplified in Figs.~\ref{fig:Fig1}(c) and (d) where we present the spatial current pattern and $I_L$, respectively, for $\mu_c=2.4t$.

\begin{figure}
\includegraphics[width=8.cm]{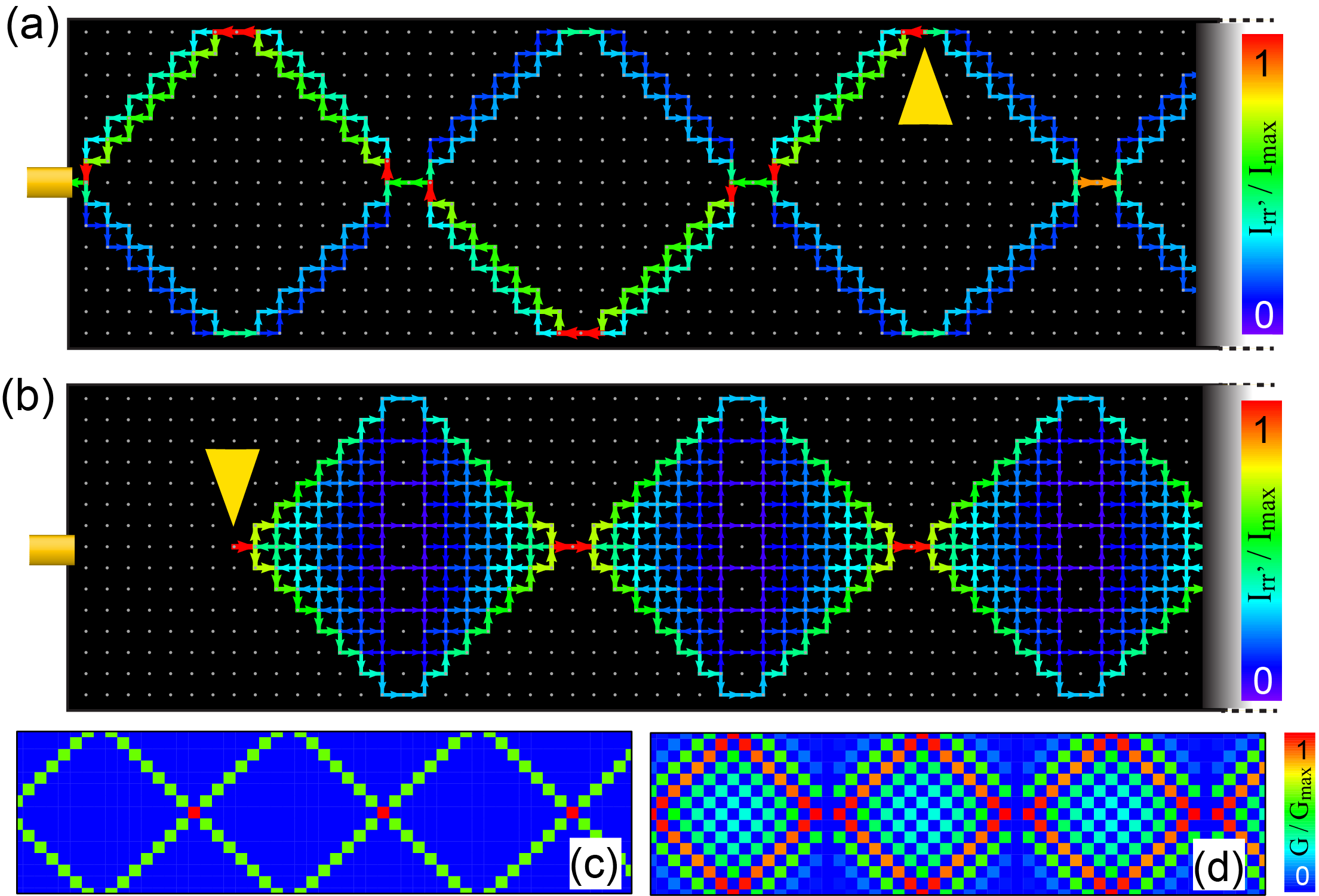}
\caption{(color online) (a),(b) Spatial pattern of current flowing from the STM tip to the leads for two different tip positions (indicated by yellow triangles) and $\mu_{T} = 5\times 10^{-5} t$, $\mu_c^T=0$. (c) Re$G^{r}_{{\bf r,L}}$ and (d)${\rm Im} G^{r}_{{\bf r,L}}$ for the case of Fig.~\ref{fig:Fig1}(a). } \label{fig:Fig1a}
\end{figure}

A physical understanding of the good agreement between $I_{{\bf r, r}^\prime}$ and $I_L$ can be gained by considering how the current flows from the STM tip through the network into the left lead. Consider, for example, the case when the STM tip is located above a network site that is part of the coherent ``current riverbed" of Fig.~\ref{fig:Fig1}(a). In this case, as shown in Fig.~\ref{fig:Fig1a}(a), the current injected from the STM tip utilizes the current riverbed, resulting in a large $I_L$ flowing from the tip to the left lead. In contrast, when the tip is positioned away from the riverbed, as shown in Fig.~\ref{fig:Fig1a}(b), the current injected by the STM tip cannot flow to the left lead, but only to the wide right lead (at infinity), since only the latter can be connected to the tip location by a diagonal path. While the spatial form of current riverbeds varies with the position of the leads and the gating of the network, this physical explanation for the success of the imaging method holds for all cases we have considered.
A more technical understanding of the good
agreement between $I_{{\bf r, r}^\prime}$ and $I_L$, is obtained by noting that their analytical expressions in Eqs.(\ref{eq:Ianalytic}) and (\ref{eq:IL}) are determined by (different) combinations of the real and imaginary parts of the non-local Green's function, $G^r_{{\bf r,L}}$. A spatial plot of these two quantities shown in Figs.~\ref{fig:Fig1a}(c) and (d) [for the case of Fig.~\ref{fig:Fig1}(a)] reveals that their spatial structures are similar on the atomic scale.
As a result, $I_L$ [see Eq.(\ref{eq:IL})] is strongly peaked in those regions where $I_{{\bf r, r}^\prime} \not = 0$ (the latter being predominantly determined by ${\rm Re} G^{r}_{{\bf r,L}}$ \cite{Can12}), thus explaining the good spatial agreement between $I_L$ and $I_{{\bf r, r}^\prime}$. The same conclusion also holds for the case of $\mu_c=2.4t$ in Figs.~\ref{fig:Fig1}(c) and (d). As follows from Eqs.(\ref{eq:Ianalytic}) and (\ref{eq:IL}), and confirmed by an extensive survey of networks, a spatial structure of ${\rm Re} G^{r}_{{\bf r,L}}$ and ${\rm Im} G^{r}_{{\bf r,L}}$ that is similar on the atomic scale is a sufficient condition for $I_L$ to provide an atomically resolved image of current paths (see discussion below).

\begin{figure}
\includegraphics[width=8.cm]{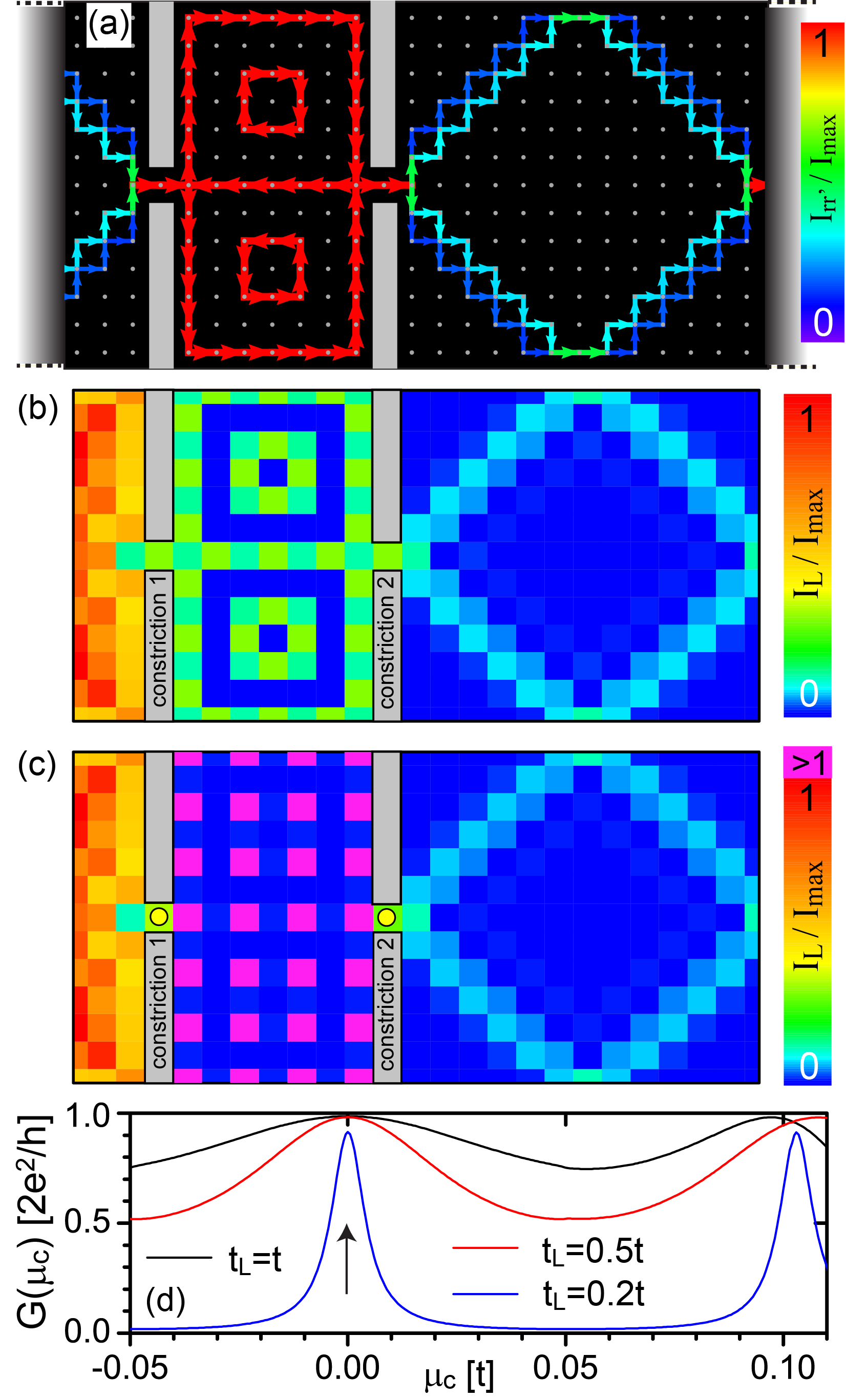}
\caption{(color online) (a) Spatial current pattern and (b) $I_L({\bf T})$ for $\mu_c=0$. (c) $I_L({\bf T})$ for $t_L=0.2$. The scales in (b) and (c) are the same. Constrictions are shown in gray. (d) $G(\mu_c)$ for several $t_L$.} \label{fig:Fig4a}
\end{figure}

The STM method also images more complex current patterns, which occur, for example, in systems with multiple constrictions, as shown for an infinitely long wire with $N_y=13$ rows in Fig.~\ref{fig:Fig4a}(a).
%
%
While in the left and right regions of this network, the current follows a diagonal path, a more complex current pattern exhibiting circulating current loops \cite{Can12} exists in the central region.
As shown in Fig.~\ref{fig:Fig4a}(b), $I_L$ correctly images the current pattern in the central and right regions, but reflects the spatial form of the density of states at $\mu^T_c$ in the left region. Conversely, $I_R$ (not shown) correctly images the spatial current pattern only in the left and central regions. To understand why $I_{L,R}$ reproduce the current pattern only in parts of the system (while together, $I_{L,R}$ yield a complete image of the current paths), we note that when the tip is located to the right of constriction 1 [Fig.~\ref{fig:Fig4a}(b)], the current flowing from the tip into the left lead needs to pass through the narrow opening in constriction 1, whose role is therefore similar to that of the narrow left lead in Fig.~\ref{fig:Fig1}(a). As a result, $I_L$ is large whenever the tip is located above the current riverbed of Fig.~\ref{fig:Fig4a}(a).
In contrast, in the region left of constriction 1, the tip, independent of its location, can always be connected to the wide left lead (at infinity) by a diagonal path, and  $I_L \approx I_{tip}$ therefore images the density of states. We thus conclude that in order for $I_L$ to correctly image the current pattern in a given region, the number of allowed current channels need to be restricted, either through the presence of a constriction, as in Fig.~\ref{fig:Fig4a}(b), or of a narrow lead, as in Fig.~\ref{fig:Fig1}(c).
%
%
\begin{figure}
\includegraphics[width=8.cm]{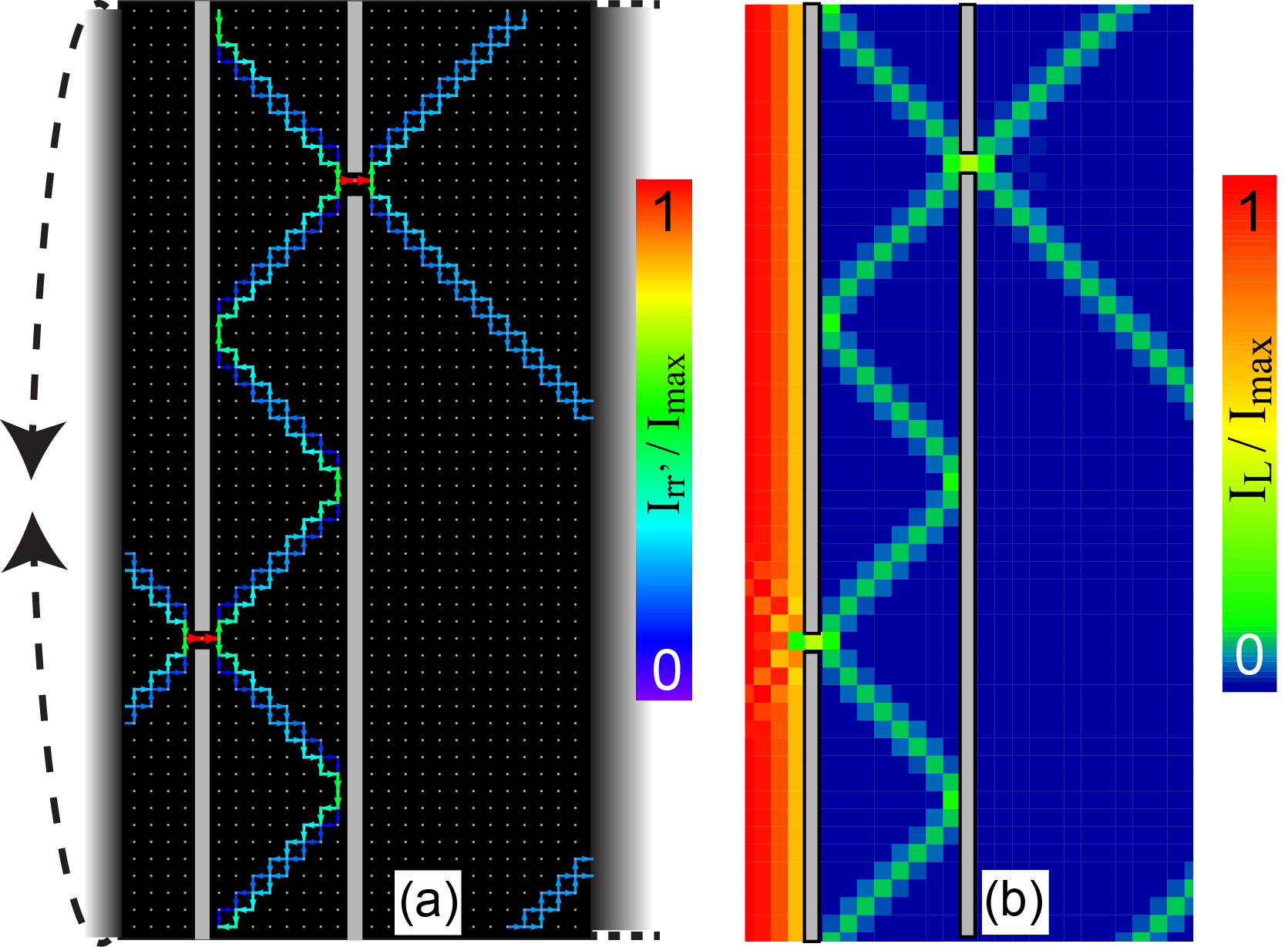}
\caption{(color online) (a) Spatial current pattern for $\mu_c=0$, and (b) $I_L({\bf T})$ for a cylinder (cut along the axis and flattened out). Constrictions are shown in gray.} \label{fig:Fig4b}
\end{figure}
Similar conclusions also hold for more complex network topologies, such as the infinitely long cylinder of circumference $N_y=54$ shown in Fig.~\ref{fig:Fig4b}(a) possessing two constrictions rotated by a relative angle of $\pi$ around the cylinder axis. Here, $I_L$ [see Fig.~\ref{fig:Fig4b}(b)] correctly images the winding current pattern only in the central and right cylinder regions, while $I_R$ provides the complementary current image.

To make the imaging method a versatile tool in the study of nanoscopic networks, it is necessary to identify an experimentally verifiable criterion for the agreement (or the lack thereof) between $I_L$ and the current pattern. Such a criterion is provided by the network's overall conductance, $G(\mu_c)$, as can be exemplified by varying the hopping, $t_{L}$, between sites in constrictions 1 and 2 [see yellow dots in Fig.~\ref{fig:Fig4a}(c)] and the central region of the network shown in Fig.~\ref{fig:Fig4a}. As discussed above, for $t_{L}=t$, the STM method provides a correct spatial image of the current pattern while at the same time, the conductance around $\mu_c=0$ [see arrow in Fig.~\ref{fig:Fig4a}(d)] is close to the (maximal) quantum of conductance, varying only weakly with $\mu_c$. However, with decreasing $t_L$, the agreement between the STM image and the actual current pattern (the latter remaining qualitatively unchanged) worsens in the central region and is completely lost around $t_L=0.2t$ [see Fig.~\ref{fig:Fig4a}(c)]. This break down is accompanied by a qualitative change in $G(\mu_c)$: it becomes sharply peaked around $\mu_c=0$, and its overall magnitude decreases. Both changes possess the same origin: for $t_L \rightarrow 0$, only a single state contributes to the charge transport at $\mu_c=0$ in the central region, implying that $G(\mu_c)$ becomes sharply peaked while the overall magnitude of ${\rm Im}G_{{\bf r} , {\bf L}}^{r}(\mu_{c}) $ scaling with $\delta ^{-1}$ becomes much larger than that of ${\rm Re} G_{{\bf r} , {\bf L}}^{r}(\mu_{c})$. As a result, $I_L$ is dominated by ${\rm Im}G$ and therefore does not correctly image the current pattern, whose spatial form is in general determined by ${\rm Re} G$ \cite{Can12}. We find that in general, the onset for the break down of the imaging method occurs when the widths of the peaks in $G(\mu_c)$ become less than the separation between peaks [see $t_L=0.5t$ in Fig.~\ref{fig:Fig4a}(d)]. In contrast, even for $t_L=0.2t$, $I_L$ still correctly images the current pattern in the right region since due to its semi-infinite nature, ${\rm Im}G$ does not scale with $\delta ^{-1}$ for $t_L \rightarrow 0$, and the overall magnitude of ${\rm Im}G$ and ${\rm Re} G$ remain the same (though the overall magnitude of $I_L$ rapidly decreases for $t_L \lesssim 0.2t$). This relation between the success of the imaging method and $G(\mu_c)$ holds for all networks we have considered: whenever $G(\mu_c)$ is close to the maximal allowed conductance, and varies weakly around a given $\mu_c$, the STM method yields a correct current image in the entire network. On the other hand, if $G(\mu_c)$ is sharply peaked (or reduced from its maximal value due to dephasing, as discussed below), the agreement breaks down in at least part of the network. Thus, $G(\mu_c)$ provides an experimentally verifiable criterion for the success of the imaging technique.

To investigate how the ability of $I_{L,R}$ to image current patterns is affected by interactions and the resulting dephasing, we consider the effects of the electron-phonon interaction in Eq.(\ref{eq:Hamiltonian})\cite{Gal07,Mon02}. The full solution of the self-consistent Dyson-equation for ${\hat G^{K,r}}$ for general $g$, $\omega_0$ and temperature $T$ is currently beyond our computational abilities, and thus reserved for future work. However, in the high-temperature approximation $ k_B T \gg \omega_0$ (with $\omega_0 \rightarrow 0$) \cite{Bih05}, the fermionic self-energy greatly simplifies to $\Sigma^{K,r} = \gamma G^{K,r}$, where $\gamma =g^{2}\left[ 1 + 2 n_{B}(\omega_{0})\right]$, and the Dyson equation can be numerically solved even for larger networks \cite{Can12}.
\begin{figure}
\includegraphics[width=8.cm]{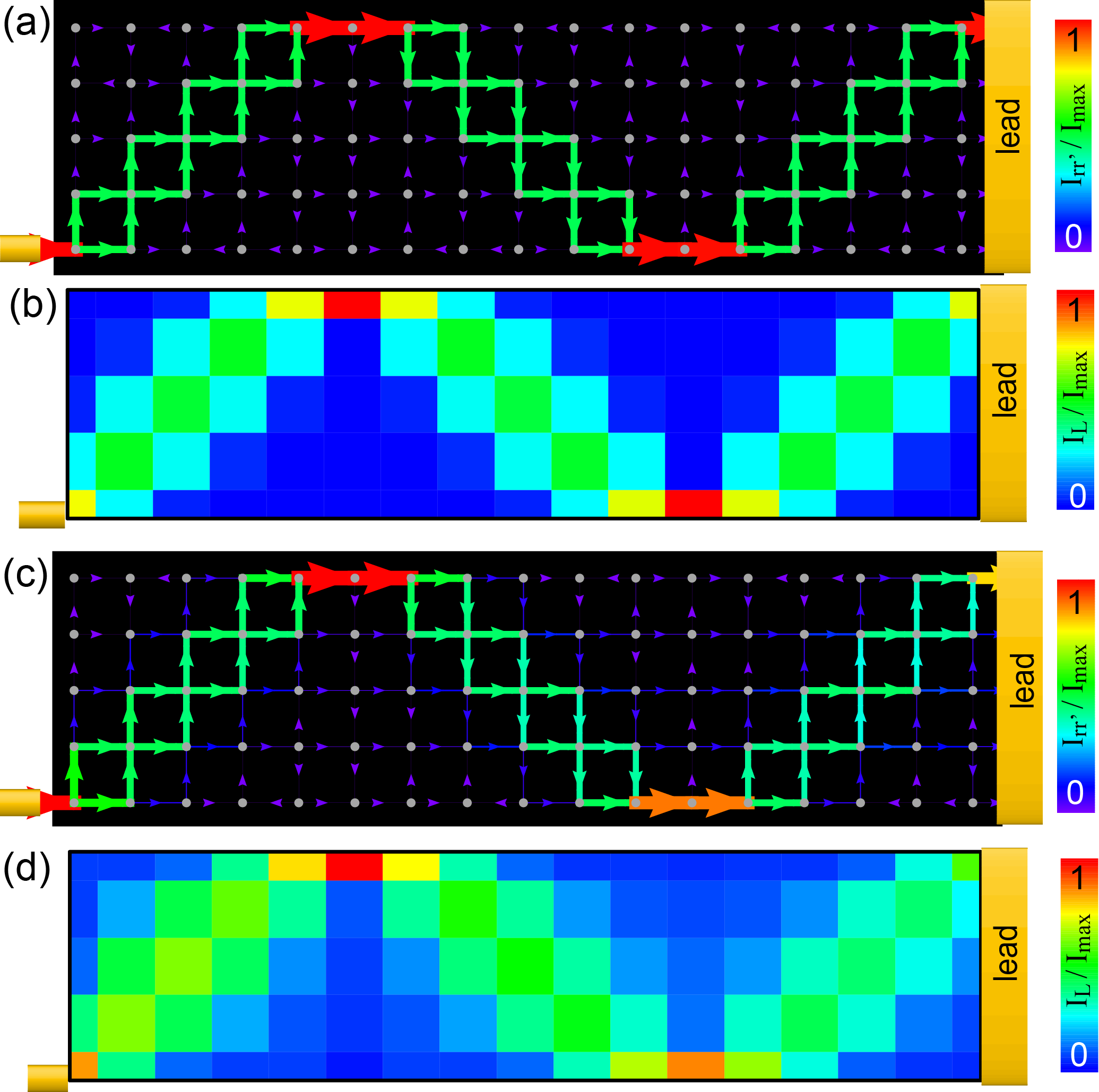}
\caption{(color online) Spatial current pattern and $I_L$ for (a),(b) $\gamma=0.002t^2$ and (c),(d) $\gamma =0.032 t^2$ with $\mu_{L} = - \mu_{R} = 0.03t$, $k_{B} T = 5\times 10^{-6} t = 10 \omega_{0}$, and $\delta = 10^{-16}t$.} \label{fig:Fig5}
\end{figure}
In Fig.~\ref{fig:Fig5} we plot the spatial current pattern and $I_L$ for a $(17 \times 5)$ network attached to a narrow and a wide lead, and two different values of the dephasing parameter $\gamma$. In general, with increasing $\gamma$, the current pattern evolves from that of the ballistic limit to that of a classical resistor network \cite{Can12}. For small $\gamma$ up to $\gamma \sim 0.002 t^{2}$ [corresponding to a mean-free path of $l \approx 13 a_0$, where $a_0$ is the lattice constant], with conductance $G\approx 0.99 \left(2 e^{2}/h\right)$, $I_L$ [Fig.~\ref{fig:Fig5}(b)] correctly images $I_{\bf r, r^\prime}$ [Fig.~\ref{fig:Fig5}(a)] which still largely resembles the current pattern of the quantum ($\gamma=0$) limit.
For larger $\gamma= 0.032 t^2$ with $l \approx 3.3 a_0$ and $G\approx 0.93 \left(2 e^{2}/h\right)$, both $I_{\bf r, r^\prime}$ [Fig.~\ref{fig:Fig5}(c)]  and $I_L$ [Fig.~\ref{fig:Fig5}(d)] become spatially more diffuse, however, $I_L({\bf T})$ still reproduces the main features of the current pattern even for such a small mean-free path. This demonstrates that the STM imaging technique is rather robust against dephasing effects even for mean-free paths as small as a few lattice spacings. Increasing $\gamma$ even further results in an overall suppression of $G(\mu_c)$, and an increasing disagreement between $I_{L}$ and $I_{\bf r, r^\prime}$, in agreement with the criterion formulated above.


We would like to thank G. Engel, P. Guyot-Sionnest, H. Jaeger, and L. Kadanoff for
stimulating discussions. This work is supported by the U.S. Department of Energy
under Award No.~DE-FG02-05ER46225 (D.K.M) and by a Department of Education GAANN Fellowship (T.C.).

\end{document}